\documentstyle[prl,aps,twocolumn,epsfig]{revtex}

\begin{document}
\date{\today}
\title{\bf{Nuclear fragmentation: sampling the instabilities of binary\\
       systems}}

\author{V. Baran$^{ab)}$, M. Colonna$^{a)}$, M. Di Toro$^{a)}$,
V. Greco$^{a)}$}

\address{$^{a)}$ Laboratori Nazionali del Sud, Via S. Sofia 44,
I-95123 Catania, Italy and University of Catania \\
$^{b)}$ NIPNE-HH, Bucharest, Romania}
\maketitle

\begin{abstract}
We derive 
stability conditions of Asymmetric Nuclear Matter ($ANM$)
and discuss the relation to mechanical and chemical instabilities
of general two-component systems. We show that the chemical instability may
appear as an instability of the system against isoscalar-like rather than
isovector-like fluctuations if the interaction between the two constituent
species has an attractive character as in the case of $ANM$. 
This leads to a new kind of liquid-gas phase transition,
of interest for fragmentation experiments with radioactive beams.
\end{abstract}

\vspace{0.5cm}

\hspace{-\parindent}PACS numbers: 21.65.+f, 25.70.Pq, 71.10.Ay

A binary system manifests a richer thermodynamical behaviour
as a consequence of a new parameter, the concentration, which is
required for a complete description of its states.
The phase transitions are more complex because they have to 
accommodate one more conservation law. A thermodynamical
state can be unstable not only mechanically but also chemically.

Nuclear matter belongs to this class of systems.
The process of multifragmentation following the collision of heavy nuclei
at medium energies is expected to show
features analogous to
usual liquid-gas phase transitions of water \cite{sau76,jaq84,bersim83}. 
In particular, the spinodal decomposition appears to be an important 
mechanism leading 
to phase separation in symmetric nuclear matter
($SNM$) 
\cite{bersim83,peth87,heis88,mar1}, as confirmed in recent experiments
\cite{bbord00}.
The relevance of instabilities in $ANM$-fragmentation has been already 
discussed few years
ago \cite{serot,bao97,bar98}. It was shown that in these systems a kind of
diffusive (or chemical) spinodal rather than the mechanical spinodal is
significant.
However the detailed nature of fluctuations responsible for such instability
has not been clarified yet. This is a quite important aspect since,
as we will see, it provides reliable information on the character of the
interaction in the medium. We remark that the effect is 
leading to the "isospin distillation" that can be experimentally
observed, see the recent data \cite{xu00}. 

In one-component systems the mechanical instability is related to 
instability against density fluctuations as the result of the strong
attraction between constituents. In symmetric binary systems,
like $SNM$, one encounters two kinds of density fluctuations:
i) isoscalar, when the densities
of the two components oscillate in phase with equal amplitude,
ii)  isovector when the two densities fluctuate still with equal
amplitude but out of phase. Mechanical instability is associated with
instability against isoscalar fluctuations leading to cluster
formation while chemical
instability is related to instability against isovector fluctuations,
leading to species separation.
We will show that in asymmetric binary systems, as in $ANM$,
this direct correspondence between the nature of fluctuations and the
thermodynamical instability is lost.

An appropriate framework for this investigation is provided by the Fermi
liquid theory which has been already applied to the study of instabilities
in symmetric binary systems as $SNM$ (the two components being protons and
neutrons) and the liquid $^3He$ (spin-up and spin-down components)
\cite{baym78,peth88}.

We first investigate the thermodynamical stability of
$ANM$ at $T=0$ extending to the asymmetric case the formalism introduced in
\cite{baym78}. The distribution functions for protons and neutrons are:  
\begin{equation}
f_q^{(0)}(\epsilon_p^q) = 
\Theta(\mu_q - \epsilon_p^q)~, ~~~~~q=n,p         \label{Fd0}
\end{equation}
where $\mu_{q}$ are the corresponding chemical potentials.
The nucleon interaction is characterized by the Landau parameters:
\begin{equation}
F^{q_1 q_2} = N_{q_1} V^2 
\frac{\delta^2 {\cal H}}{\delta f_{q_1} \delta f_{q_2}}
~,~~N_q(T) = \int\,{-2~d{\bf p} \over (2\pi\hbar)^3} 
        {\partial f_q(T) \over \partial\epsilon_p^q}~  
           \label{ld}
\end{equation}
where $\cal H$ is the energy density, V is the volume and $N_{q}$ is the
single-particle level density at the Fermi energy.
At $T=0$ this reduces to
$N_q(0) = mp_{F,q}/(\pi^2\hbar^3)$, were $p_{F,q}$ is the Fermi momentum
of the $q$-component. Thermodynamical stability for $T=0$ requires the 
ground state energy
to be an absolute minimum for the undistorted distribution functions, 
such that the relation: 
\begin{equation}
\delta{\cal H} - \mu_p \delta \rho_p - \mu_n \delta \rho_n > 0~~ \label{varia}
\end{equation}
is satisfied when we deform proton and neutron Fermi seas.
The distorted distribution functions can be written as 
$f_q({\bf p}) = \Theta(\epsilon_{F,q}(\theta)  - \epsilon_p^q)~$,
where $\epsilon_{F,q}(\theta)$ is a direction dependent Fermi energy
characterizing the distortion. We will follow the usual multipole
expansion for the variation
\begin{equation}
[\delta \epsilon_{F,q}(\theta)] = \epsilon_{F,q}(\theta)-\mu_q=
 \sum \nu_q^l P_l(cos\theta)
\end{equation}
and for the Landau parameters
\begin{equation}
F^{q_1 q_2}(\theta_1-\theta_2) = \sum_{l}
F_l^{q_1 q_2} P_l(cos(\theta_1-\theta_2)).
\label{expa2}
\end{equation}
We limit ourselves to monopolar deformations considering here
momentum independent interactions such that
$F^{q_1q_2}_{l=0}$ are the only non-zero Landau parameters \cite{foot1}.
Then, up to second order in the variations, Eq.(\ref{varia}) becomes
\begin{equation}
\delta{\cal H} - \mu_p \delta \rho_p - \mu_n \delta \rho_n = \frac{1}{2}
(a {\nu_p}^{2} + b {\nu_n}^{2} + c \nu_p \nu_n)  \label{varia1}
\end{equation}
where $\nu_{n,p} \equiv \nu_{n,p}^0$ and
\begin{eqnarray}
a = N_p(0)(1 + F_{0}^{pp})~~;~~
b = N_n(0)(1 + F_{0}^{nn})~~; \nonumber \\ 
c = N_n(0) F_{0}^{pn} + N_p(0) F_{0}^{np} = 
2 N_n(0) F_{0}^{pn}.~~  \label{abc}
\end{eqnarray}
We diagonalize the r.h.s. of Eq.(\ref{varia1}) by introducing the following
transformation:
\begin{eqnarray}
u &=& cos\beta~ \nu_p + sin\beta~ \nu_n,    \nonumber \\
v &=& - sin\beta~ \nu_p + cos\beta~ \nu_n,         \label{rot}
\end{eqnarray}
where the {\it mixing} angle $0 \le \beta \le \pi/2$ is defined by
\begin{equation}
tg~ 2\beta = \frac{c}{a-b} = \frac{N_n(0) F_{0}^{pn} + N_p(0) F_{0}^{np}}
{N_p(0)(1 + F_{0}^{pp}) - N_n(0)(1 + F_{0}^{nn})}. \label{beta}
\end{equation}
Then Eq.(\ref{varia1}) takes the form
\begin{equation}
\delta{\cal H} - \mu_p \delta \rho_p - \mu_n \delta \rho_n =
X u^2 + Y v^{2} \label{varia2}
\end{equation}
where
\begin{eqnarray}
X &=& \frac{1}{2} (~~ a + b + sign(c) \sqrt{(a-b)^{2} + c^{2}}~~) \nonumber \\ 
&\equiv& \frac{N_p(0)+N_n(0)}{2} (~ 1 + F_{0g}^s~) \label{A}
\end{eqnarray}
and
\begin{eqnarray}
Y &=& \frac{1}{2} (~~ a + b - sign(c) \sqrt{(a-b)^{2} + c^{2}}~~) \nonumber \\
&\equiv& \frac{N_p(0)+N_n(0)}{2} (~ 1 + F_{0g}^a ~). \label{B}
\end{eqnarray}
Thanks to the rotation Eq.(\ref{rot}) we separate the total variation
Eq.(\ref{varia}) into two independent contributions, the "normal" modes,
characterized by the "mixing angle" $\beta$, which depends on the density of
states and  the details of the interaction.

In the symmetric case, $N_p=N_n \equiv N$, $F_0^{nn}=F_0^{pp}$ 
and $F_0^{np}=F_0^{pn}$,
 Eq.(\ref{varia1}) reduces to
\begin{eqnarray}
\delta{\cal H} - \mu_p \delta \rho_p - \mu_n \delta \rho_n &=& 
\frac{N(0)}{2}
(1 + F_0^s)(\nu_p + \nu_n)^{2}    \nonumber \\
 &+& \frac{N(0)}{2}(1 + F_0^a) (\nu_p -
\nu_n)^{2} \label{varia2s}
\end{eqnarray}
where $F_0^s \equiv F_0^{nn} + F_0^{np}$ and 
$F_0^a \equiv F_0^{nn} - F_0^{np}$ are symmetric and antisymmetric
(or isoscalar and isovector) Landau parameters, and we recover the usual 
Pomeranchuk stability conditions for pure isoscalar/isovector
fluctuations \cite{bar98}.

In the general case
we interpret 
$u$- and $v$-variations as new independent
${\it isoscalar}$-like and ${\it isovector}$-like fluctuations appropriate
for asymmetric systems. The proton and neutron densities will
fluctuate in phase for  isoscalar-like variations and out of
phase for isovector-like variations, see Eq.(\ref{rot}).
Moreover, $F_{0g}^s$ and $F_{0g}^a$, defined by Eq. (\ref{A},\ref{B}),
can be considered as
generalized symmetric and antisymmetric Landau parameters.

From Eq.(\ref{varia2}) we see that thermodynamical stability requires
$X>0$ {\it and} $Y>0$. Equivalently, the following conditions have 
to be fulfilled:
\begin{equation}
1 + F_{0g}^s > 0~~~~ and~~~~1 + F_{0g}^a > 0,          \label{pomegen}
\end{equation}
They represent Pomeranchuk stability conditions extended
to asymmetric binary systems.

The new stability conditions, Eq.(\ref{pomegen}),
are equivalent to mechanical and chemical stability of a 
thermodynamical state, \cite{landau}, i.e.
\begin{equation}
\left({\partial P \over \partial \rho}\right)_{T,y} > 0~~~and~~~
\left({\partial\mu_p \over \partial y}\right)_{T,P} > 0
\end{equation} 
where $P$ is the pressure and $y$ the proton fraction, as 
it can be proved by observing that \cite{foot2}:
\begin{eqnarray}
X Y &=& N_p(0) N_n(0)[(1 + F_0^{nn})(1 + F_0^{pp}) - F_0^{np}F_0^{pn}]~~~~~~
\nonumber \\
~~~~~~~ \nonumber \\
&=&\frac{(N_p(0) N_n(0))^{2}}{(1-y) \rho^{2}} 
\left({\partial P \over \partial \rho}\right)_{T,y}
\left({\partial\mu_p \over \partial y}\right)_{T,P} \label{chimec} 
\end{eqnarray}
and: 
\begin{eqnarray}
\left({\partial P \over \partial \rho}\right)_{T,y} =
\frac{\rho y (1-y)}{N_p(0) N_n(0)}(~t~ a + \frac{1}{t}~ b + c~)~~~~~~
\nonumber \\
\propto X(\sqrt{t}cos\beta + \frac{1}{\sqrt{t}}
sin\beta)^2 + Y(\sqrt{t} sin\beta - \frac{1}{\sqrt{t}} cos\beta)^{2}
\nonumber \\~~~~with~~~~
t = \frac{y}{1-y} \frac{N_n(0)}{N_p(0)}.~~~~~~~~~~
\label{compres}
\end{eqnarray}

From Eq.s(\ref{varia2},\ref{pomegen}), we are led to define as 
isoscalar instability the case when the
state is unstable against isoscalar-like fluctuations, i.e. when 
$1 + F_{0g}^s < 0$
(or $X<0$). Analogously we deal with isovector instability when the system
is unstable against isovector-like fluctuations i.e.
when $1 + F_{0g}^a < 0$ (or $Y<0$).

One can easily see that in symmetric nuclear matter
the isoscalar instability, ($X<0,Y>0$), appears as mechanical instability 
and the isovector 
instability, ($X>0,Y<0$), as chemical instability.
Indeed now
$t=1$, $a=b$, $\beta=\pi/4$ and so 
$1 + F_{0g}^s$ (or $X$) and
$\left({\partial P \over \partial \rho}\right)_{T,y}$  
are proportional (see Eq.(\ref{compres})).
This is not true any longer for an asymmetric system.

To simplify the analysis let us assume
for the moment that the quantities $a$ and $b$ remain positive
in the density range we discuss here and so we can study the effect of the
interaction between the two components, given by $c$, on the
instabilities of the mixture.
This is indeed the case of nuclear matter. 

If $c<0$, i.e. for an attractive interaction between the two components,
from Eq.(\ref{B}) we see that the system is stable against isovector-like
fluctuations but  it becomes isoscalar
unstable if $c<-2\sqrt{ab}$ (see Eq.(\ref{A})). However
thermodynamically this instability against isoscalar-like fluctuations
will show up as a chemical instability if
$(-t a -b/t ) < c < -2\sqrt{ab}$ or as a
mechanical instability if $c < (-t a -b/t ) < -2\sqrt{ab}$
(see Eq.(\ref{compres})).
{\it Therefore isoscalar-like instabilities are not necessarily 
equivalent to mechanical instabilities
and may manifest instead as chemical instabilities.}

If $c>0$, i.e. when the interaction between the components is repulsive, the
thermodynamical state is always stable against isoscalar-like fluctuation,
but can be isovector unstable if $c > 2\sqrt{ab}$. Since
the system is mechanically stable ($a,b,c>0$, see Eq. (\ref{compres})), the
isovector instability is now associated with chemical instability.
Such situation will lead to a component separation of the
liquid mixture. 

Following this line a complete analysis of the instabilities of any
binary system can be performed, in connection to signs, strengths and
density dependence of the interactions.

We show now a quantitative calculation for asymmetric
nuclear matter which illustrates 
the previous general discussion.  
We investigate the instabilities of $ANM$ characterized 
by a potential energy density of Skyrme-like type, \cite{bar98} 
\begin{eqnarray}
{\cal H}_{pot}(\rho_n,\rho_p) &=&
{A \over 2}{(\rho_n+\rho_p)^2 \over \rho_0} +
{B \over \alpha + 2}{(\rho_n+\rho_p)^{\alpha + 2} \over \rho_0^{\alpha + 1}}
\nonumber \\
 &+& (C_1 - C_2(\frac{\rho}{\rho_0})^{\alpha})
 {(\rho_n-\rho_p)^2 \over \rho_0}
                         \label{hpot}
\end{eqnarray}
where $\rho_0 = 0.16~\mbox{fm}^{-3}$ is the 
nuclear saturation density.

The values of the parameters $A=-356.8$ MeV, $B=303.9$ MeV, $\alpha=1/6$,
 $C_1=125$ MeV, $C_2=93.5$ MeV are adjusted to reproduce the saturation 
properties of symmetric nuclear matter and the symmetry energy coefficient.
We also extend the discussion to finite temperatures \cite{peth93}.

We first focus on the low density region, where phase transitions 
of liquid-gas type are expected to be seen in fragmentation events.
Since $a,b > 0$ and $c<0$, in the case of the
liquid-gas phase transition in asymmetric nuclear matter we deal only with
instability against isoscalar-like fluctuations as it happens 
for symmetric nuclear matter.
However, at variance with the latter case, now the instability can
 manifest either as 
chemical or mechanical depending on the relative strength of
the interactions in the system, as discussed previously.

\begin{figure}[htb]
\epsfysize=4.cm
\centerline{\epsfbox{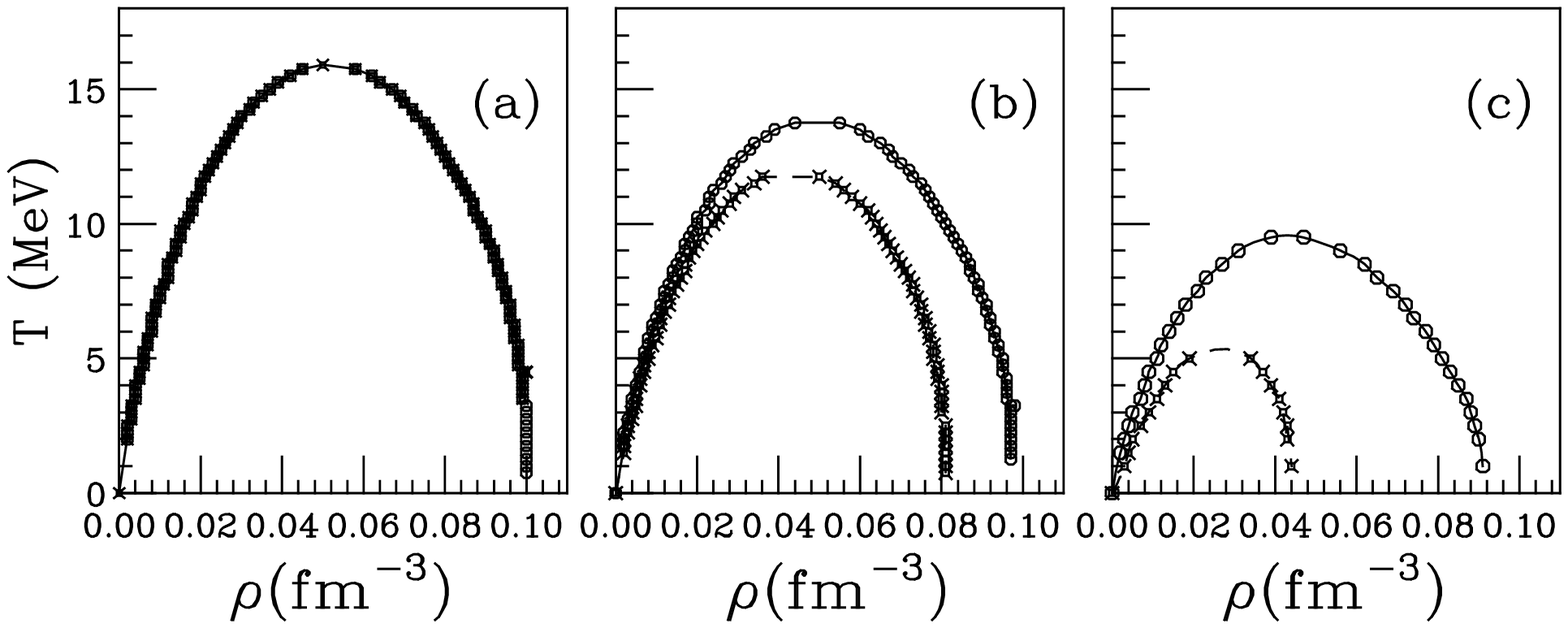}}
\caption{
Spinodal line corresponding to isoscalar-like instability of asymmetric 
nuclear matter (circles) and mechanical instability (crosses) for three
proton fractions: $y=0.5$ (a), $y=0.25$ (b), $y=0.1$ (c).
}
\end{figure}
 
In Figure 1 the circles represent the spinodal line corresponding to
isoscalar-like instability, as defined above, for three values of 
the proton fraction. For $y < 0.5$ under this border one
encounters either chemical instability, in the 
region between the two lines, 
or mechanical instability, under the inner line (crosses). The latter is
defined by the set of values ($\rho,T$) for which
$\left({\partial P \over \partial \rho}\right)_{T,y} = 0$.
We observe that the line defining
chemical instability is more robust against the variation of the proton
fraction in comparison to that defining
mechanical instability: reducing the proton fraction it becomes 
energetically less and less favoured to break in clusters with
the same initial asymmetry.
%

From Eq.s(\ref{chimec}) and (\ref{compres}), we see that the quantity
$\left({\partial\mu_p \over \partial y}\right)_{T,P}$ changes
the sign passing through zero when we cross the spinodal line corresponding to
chemical instability and passing through infinity when we cross the line
associated with mechanical instability. However, from Eq.(\ref{chimec})
(see also Fig. 3), we get 
that at the inner spinodal line, the ratio
$\left({\partial\mu_p \over \partial y}\right)_{T,P}/
\left({\partial \rho \over \partial P}\right)_{T,y} \propto XY$
is a finite negative number. Therefore we conclude that the
singularities of two quantities cancel by taking the ratio.
Note also the smooth behaviour of the mixing angle $\beta$
through the mechanical instability line, Fig.2a.

\begin{figure}[htb]
\epsfysize=4.0cm
\centerline{\epsfbox{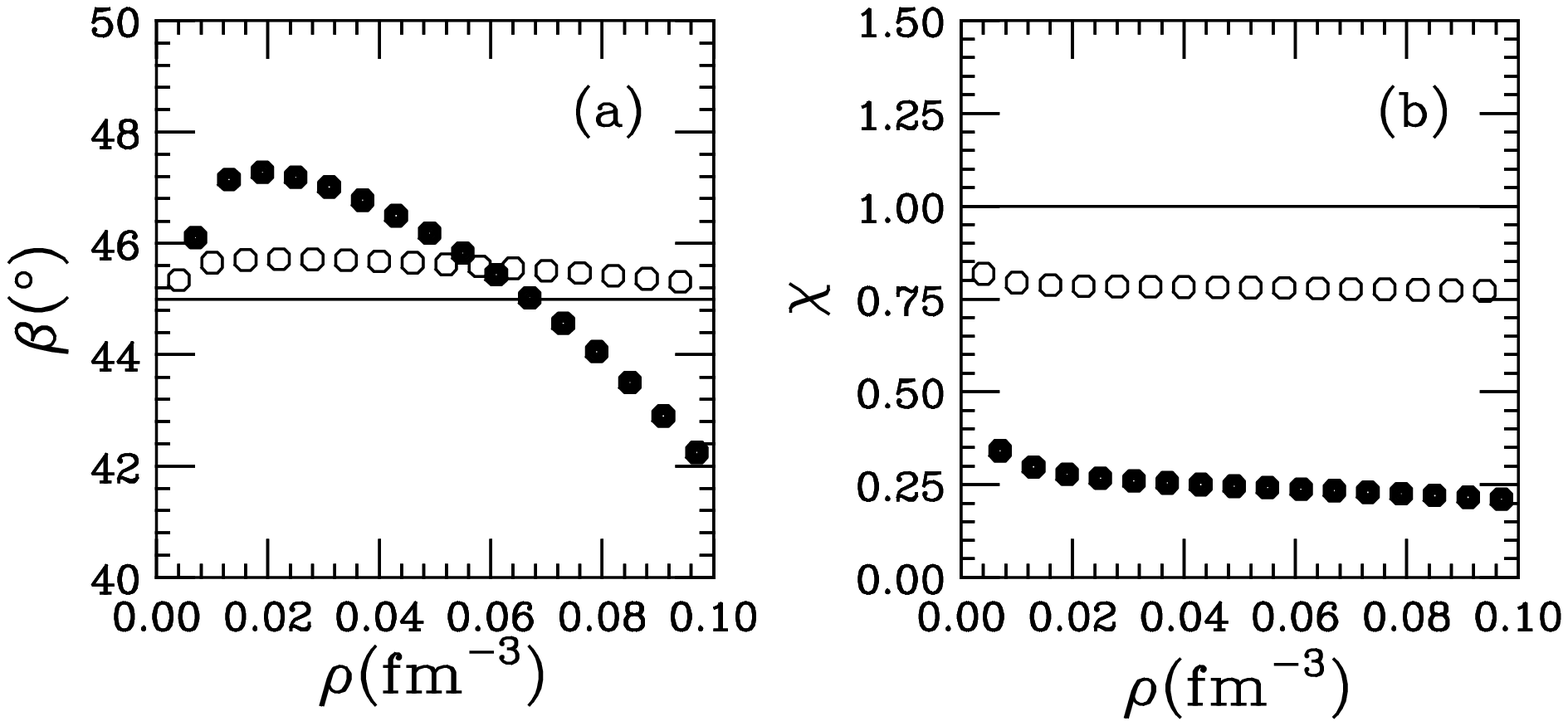}}
\caption{
Density dependence of the mixing angle $\beta$, Eq.(\ref{beta}), (a) and
of the function $\chi$ (b) 
for three proton fractions, $y=0.5$ (solid), $y=0.4$ (open circles),
and $y=0.1$ (full circles) at T=1MeV.
}
\end{figure}

Finally let us stress that, for $ANM$, even if we deal
with instabilities against isoscalar-like fluctuations, we encounter a
chemical effect, which gives raise to isospin distillation in equilibrium
phase transitions \cite{xu00}.
Indeed the variation of the asymmetry $(I=1-2y)$ is
\begin{equation}
\delta I = \nu_p [\frac{(1-I_0)N_n}{(1+I_0)N_p} tg\beta - 1] \equiv
\nu_p [\chi-1]
\end{equation}
where $I_0$ is the initial asymmetry.
For $y=0.5$ we have $\delta I=0$ but for $y<0.5$,
we find $\chi<1$  (see Fig.2b).
Therefore $\delta I < 0$ if $\nu_p>0$ and the
opposite for $\nu_p<0$.
 
To be more general,
we have extended our investigation also to the high density region. In Fig.3
we plot the density dependence of the generalized Landau parameters. We 
find that for the considered interaction the system exhibits another
instability at high density, around $1.5 fm^{-3}$, where the quantity
$c$ becomes positive and 
$1 + F_{0g}^a < 0$ (or $Y<0$).

\begin{figure}[htb]
\epsfysize=4.0cm
\centerline{\epsfbox{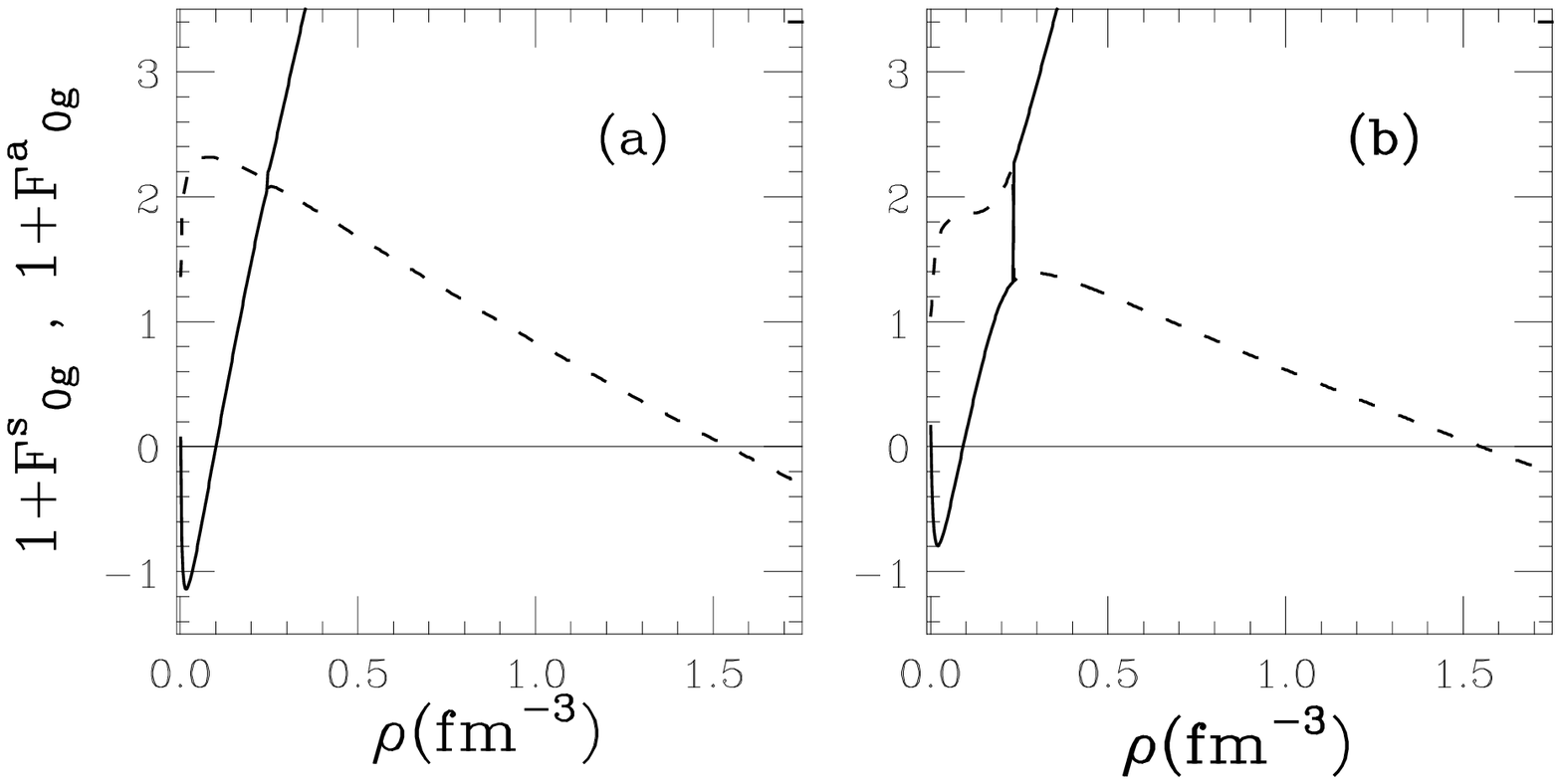}}
\caption{
Density dependence of the generalized Landau parameters for two proton
fraction, $y=0.4$ (a) and $y=0.1$ (b) (symmetric, solid and 
antisymmetric, dashed)
at T=1MeV.
}
\end{figure}

Thermodynamically we see from Eq.(\ref{chimec}) that this is again a chemical
instability. However now it
results from isovector-like fluctuations, in contrast to the low density
instability. The
reason is the change in the character of the interaction between
the two components. Since the interaction becomes repulsive the nuclear phase
can become unstable against proton-neutron separation.
We also notice in Fig.3 that the generalized Landau parameters display 
a discontinuity where the quantity $c$ changes the sign.
Other effective forces, with more repulsive symmetry terms, will not show
this high density chemical instability \cite{col98}, that actually could
be of interest for other many body systems.

In conclusion, we have shown that in asymmetric binary systems
the relevant instability regions are defined by the instabilities against
isoscalar-like and isovector-like fluctuations. The  kind of thermodynamical
instability, chemical or mechanical, will
depend on the relative strength between the various interactions acting in the
system.

In particular, the 
liquid-gas phase transition in asymmetric nuclear matter results
from instability against isoscalar-like modes rather than isovector-like
due to the attractive character of the interaction between protons
and neutrons. This is a qualitatively new effect which is leading
to the observed "isospin distillation" \cite{xu00}. More data are
expected from the new Radioactive Beam facilities opening the possibility
of direct studies of the charge dependent part of the nuclear interaction
far from normal conditions.


\begin{thebibliography}{99}
\bibitem{foot1} The extension of the formalism to momentum dependent
interactions is straightforward: see \cite{baym78} or \cite{col98}
for an approach based on kinetic Landau equations.
\end{thebibliography}
\end{document}